\documentclass[10pt, conference, letterpaper]{IEEEtran}
\usepackage{url}
\usepackage{amssymb}
\usepackage{amsfonts}
\usepackage{amsmath}
\usepackage{amsmath,amssymb,verbatim}
\usepackage{graphicx}
\usepackage{float}
\usepackage{epsfig}
\usepackage{subfig}
\usepackage{url}
\usepackage{threeparttable}
\usepackage{algorithm}
\usepackage{algorithmic}
\usepackage{color}
\usepackage{graphicx,dblfloatfix}
\usepackage{changepage}%
\ifCLASSINFOpdf
\else
\fi

\newtheorem{theorem}{Theorem}

\newtheorem{definition}{Definition}

\newtheorem{lemma}{Lemma}

\floatname{algorithm}{Algorithm}

\begin{document}
%

\title{Optimization Problems in Correlated Networks}
\author{Song Yang, Stojan Trajanovski and Fernando A. Kuipers\\
Delft University of Technology, The Netherlands\\\{S.Yang, S.Trajanovski, F.A.Kuipers\}@tudelft.nl}
\maketitle



%

\begin{abstract}
Solving the shortest path and the min-cut problems are key in achieving high performance and robust communication networks. Those problems have often beeny studied in deterministic and independent networks both in their original formulations as well as in several constrained variants. However, in real-world networks, link weights (e.g., delay, bandwidth, failure probability) are often correlated due to spatial or temporal reasons, and these correlated link weights together behave in a different manner and are not always additive. 

In this paper, we first propose two correlated link-weight models, namely (i) the deterministic correlated model and (ii) the (log-concave) stochastic correlated model. Subsequently, we study the shortest path problem and the min-cut problem under these two correlated models. We prove that these two problems are NP-hard under the deterministic correlated model, and even cannot be approximated to arbitrary degree in polynomial time. However, these two problems are polynomial-time solvable under the (constrained) nodal deterministic correlated model, and can be solved by convex optimization under the (log-concave) stochastic correlated model.
\end{abstract}


\numberwithin{equation}{section} \renewcommand\theequation{\arabic{section}.\arabic{equation}}

\section{Introduction}
Both the shortest path problem and the min-cut problem are of great importance to various kinds of network routing applications (e.g., in transportation networks, optical networks, etc.). A traffic request can be routed in the most efficient way (e.g., with minimum delay) by computing a shortest path. On the other hand, the min-cut problem arises in the context of network reliability, network throughput, etc. Fortunately, both of these problems are solvable in polynomial time for networks with independent additive link weights. 

However, often correlations or (inter-)dependencies exist among link weights. For example, in overlay \cite{Cui02} or multilayer networks \cite{Kuipers09}, the abstract links in the logical layer are mapped to different links in the physical layer. In this context, two or more abstract links, which use the same physical links, may have correlated latencies \cite{Savage99}, bandwidth usage \cite{Kostic03}, or geographical failures \cite{Kim10}, \cite{KuipersRegion}. Or if the path must pass through some specific nodes (e.g., regenerators to boost the signal quality \cite{KuipersICNP}), such important nodes and their links may also introduce dependencies. Correlations also appear in social networks. For instance, a message may be forwarded more rapidly if it came from a close friend rather than from a distant acquaintance. Another example relates to interdependent networks \cite{Buldyrev2010}, where for instance the electricity network and Internet are coupled and inter-connected, and one node or link failure in one network may cause failures of nodes or links in the other network. Similarly in Shared-Risk Link Group (SRLG) networks \cite{Strand01}, links in, for example, the same duct will fail simultaneously, if their duct fails. The dependencies in interdependent and SRLG networks can also be seen as correlations, so we use the term correlation throughout this paper and study relevant problems in these so-called correlated networks. Our key contributions are as follows:
\begin{itemize}
\item We propose two correlated link weight models, namely a \emph{deterministic correlated model} and a \emph{stochastic correlated model}.

\item We study the shortest path problem and the min-cut problem under the deterministic correlated model, and we prove that both of them are NP-hard and even cannot be approximated in polynomial time. 

\item On the other hand, we also show that both the shortest path problem and the min-cut problem are solvable in polynomial time under a (constrained) nodal deterministic correlated model.

\item To solve both problems under the proposed correlated models, we propose exact algorithms under the deterministic correlated model, and develop convex optimization formulations for the stochastic correlated model.

\end{itemize}

The remainder of this paper is organized as follows. Section~\ref{Sec:CorreModels} introduces our two correlated link weight models. In Section \ref{Sec:CorreRouting} and Section~\ref{Sec:CorreCut}, we study the shortest path problem and min-cut problem, respectively, for the proposed models and devise algorithms to solve them exactly. An overview of the related work is presented in Section~\ref{Sec:CorreRelWork} and we conclude in Section~\ref{Sec:CorreConclusion}.

\section{Correlated Link Weight Models} \label{Sec:CorreModels}
A network having node and link weights can be transformed into a directed network with only link weights, as done in \cite{Dantzig03}. Therefore, we assume nodes are unweighted and only consider correlated link weights. Throughout this paper, we use the term ``correlated model'' to represent ``correlated link weight model''. 

\subsection{Deterministic Correlated Model} \label{Sec:DetModel}
\begin{figure}[tbh]
\centering
\includegraphics[trim = 0mm 0mm 0mm 0mm,clip,width=0.33\textwidth]{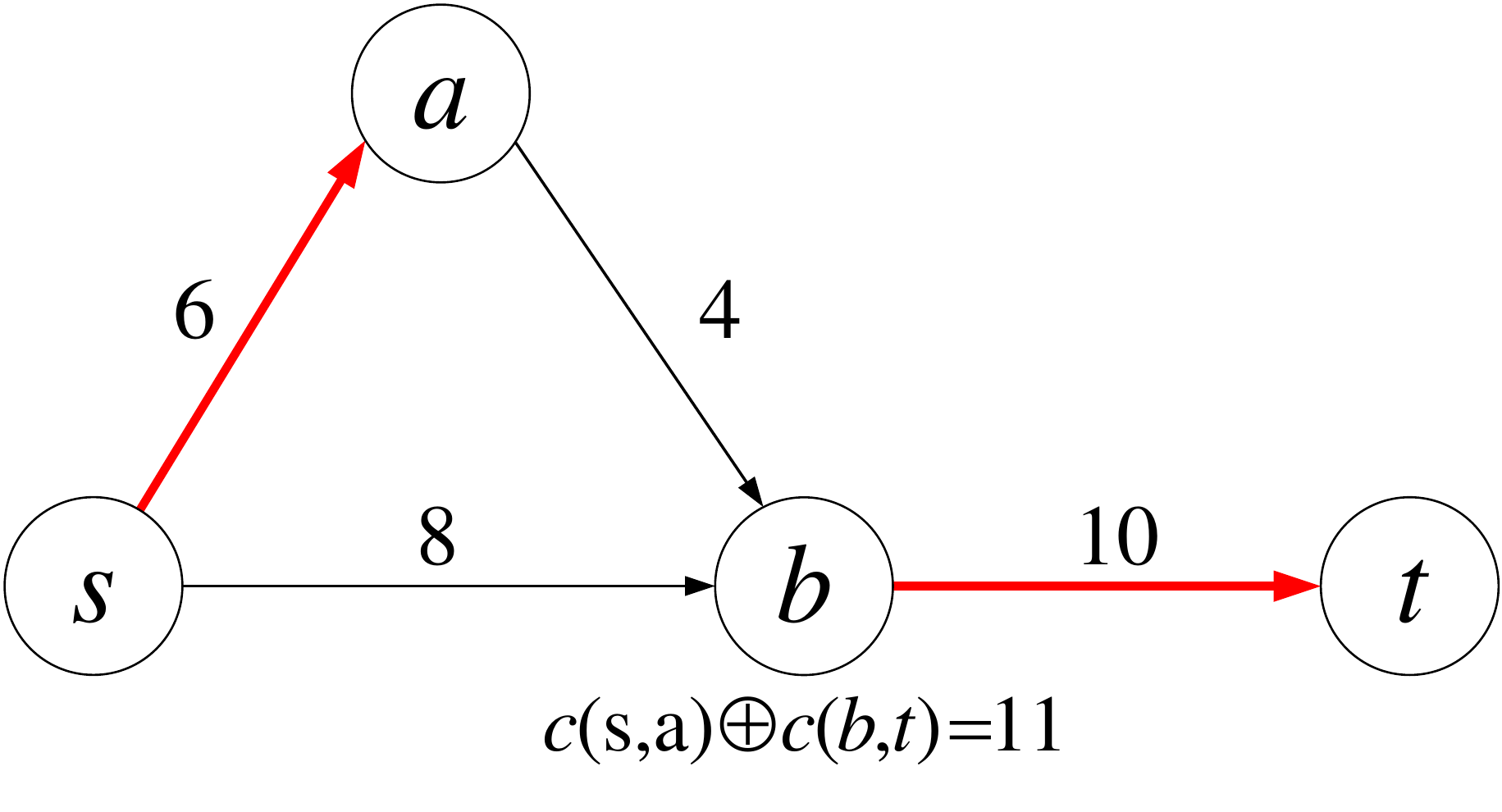}
\caption{An example of the deterministic correlated model.}%
\label{Fig: Nondom}%
\end{figure}

Without loss of generality, we use $w(l)$ to represent the weight of link $l$. For simplicity, in this paper we call $w(l)$ the cost of $l$, although it could also reflect other metrics such as delay, energy, etc. In the deterministic correlated model, for any two links $l_i$ and $l_j$, their joint total cost is represented by $w(l_i) \oplus w(l_j)$, where the operator $\oplus$ indicates the joint total cost of the links, which may differ from the $+$ operator when they are \emph{correlated}. When correlated, the use of one link may influence the cost of another in this model. For example, in Fig. \ref{Fig: Nondom} where the cost is shown above each link, it is assumed that only links $(s,a)$ and $(b,t)$ are correlated with joint cost of $11$, and all the other links have uncorrelated costs. We can see that in path $s$-$b$-$t$, the cost of link $(b,t)$ is $10$, since another link $(s,b)$ in this path is not correlated with it. Therefore this path's cost is equal to $18$. However, in path $s$-$a$-$b$-$t$, the cost of link $(b,t)$ should be calculated together with link $(s,a)$, leading to a joint cost $11$, since they are correlated and both appear in this path. Therefore, this path's total cost is equal to $11+4=15$, which is smaller than the sum of the individual link costs ($6+4+10=20$).

Equivalently, we could formulate $w(l_i) \oplus w(l_j)=\rho_{i,j} \cdot (w(l_i)+w(l_j))$, where $\rho_{i,j}$ stands for the correlation coefficient between links $l_i$ and $l_j$, and its value varies in the range of $(0,\infty)$, since we do not consider negative costs. When $\rho_{i,j}$ is equal to $1$, $l_i$ and $l_j$ are uncorrelated, when $\rho_{i,j}$ is greater than $1$, $l_i$ and $l_j$ have an increasing correlation, and otherwise we say that $l_i$ and $l_j$ have a decreasing correlation. 

Analogously, for given $m>1$ links $l_1,l_2,...,l_m$ in the deterministic correlated model, their joint total cost can be expressed as follows:
\begin{equation}
\label{Eq: SumCorrelated}
w(l_1) \oplus w(l_2) \cdot \cdot \cdot \oplus w(l_m)=\rho_{1,2,...,m} \cdot (w(l_1)+w(l_2)+\cdot \cdot \cdot+w(l_m))
\end{equation}
Similarly, if the link $l$'s weight is multiplicative (e.g., failure probability), then by using $-log(w(l))$ to represent its weight value, Eq.~(\ref{Eq: SumCorrelated}) also applies.
The decreasing correlation case can also reflect SRLG networks. 
For instance, in SRLG networks, each link is associated with several SRLG events with their respective failure probabilities. Hence, the total failure probability (represented by $P_{SRLG}$) of two correlated links that have at least one SRLG in common will be equal to the product of the failure probabilities of all the distinct SRLG events that belong to these two links. Let us denote $P_{l_1}=P_{s_1} \cdot P_{s}$ and $P_{l_2}=P_{s_2}\cdot P_{s}$ as the failure probabilities of these two links, respectively, where $P_{s}$ denotes the common SRLGs' failure probability between $l_1$ and $l_2$, and $P_{s_1}$ ($P_{s_2}$) is the non-overlapping SRLGs' failure probability of $l_1$ ($l_2$). Then $P_{l_1}\cdot P_{l_2}<P_{SRLG}=P_{s_1} \cdot P_{s_2} \cdot P_{s} <min(P_{l_1},P_{l_2})$. By taking the $-log$, we have:

\scriptsize
\begin{equation}
\max(-\log(P_1),-\log(P_2))<-\log(P_{SRLG})<(-\log(P_{l_1})+(-\log(P_{l_2})) \nonumber
\end{equation}
\normalsize
Or equivalently,
\begin{equation}
-\log(P_{SRLG})=\rho \cdot (-\log(P_{l_1})+(-\log(P_{l_2})) 
\end{equation}
where $\rho<1$ denotes their correlation coefficient. 

In probability theory, given two random variables $X$ and $Y$ with expected values $\mu_X$ and $\mu_Y$, and standard deviations $\sigma_X$ and $\sigma_Y$, their \emph{linear} correlation coefficient $\rho(X,Y)$ is defined as:
\begin{equation}
\label{Eq:PbCorreCoe}
\rho(X,Y)=\frac{Cov[X,Y]}{\sigma_X \sigma_Y}=\frac{E[(X-\mu_X)(Y-\mu_Y)]}{\sigma_X \sigma_Y}
\end{equation}
where $Cov[X,Y]$ represents the covariance of $X$ and $Y$.

However, the \emph{linear} correlation coefficient in probability theory is different from and cannot be transformed to the one defined in the deterministic correlated model, because: the variances of $X$ and $Y$ in Eq. (\ref{Eq:PbCorreCoe}) must be nonzero and finite. However, in the deterministic correlated model, for any link $l$, when none of its correlated links simultaneously appear on a path, the cost of $l$ is fixed/deterministic with variance of $0$.

\subsection{Stochastic Correlated Model}
In many real-life networks, the link weights are uncertain because of inaccurate Network State Information (NSI) \cite{Lorenz98,Guerin99}. For instance, Papagiannaki \emph{et al.} \cite{Papagiannaki02} showed that the queuing delay distribution can be approximated by a Weibull distribution. Since the Cumulative Density Function (CDF) of a Weibull distribution is log-concave and the CDFs of many common distributions (e.g., Exponential distribution, Uniform distribution, etc.) are log-concave \cite{Bagnoli07, Mohtashami11}, we make, as in \cite{Lorenz98,ICNP14}, a mild (general) assumption that the link weights follow a log-concave distribution.

We first define the \emph{Correlated Group} (CG): 
\begin{definition}
Given is a network $G(\mathcal{N},\mathcal{L})$ where $\mathcal{N}$ represents a set of $N$ nodes and $\mathcal{L}$ denotes a set of $L$ links. A Correlated Group (CG) is a subset of links $L_{CG} \subseteq \mathcal{L}$, and $\forall l \in L_{CG}$, $\exists l' \in L_{CG} \backslash \{l\}$, such that $l$ and $l'$ are correlated ($\rho_{l,l'} \neq 1 $). 
\end{definition}
Accordingly, the Maximum Correlated Group (MCG) is defined as a CG with the maximum number of correlated links. If a link $l$ is uncorrelated/independent with all the other links, then we say $\{l\}$ is a single element MCG. Suppose there are $\Omega$ Maximum Correlated Groups (MCGs), and there are $m_i>0$ links (denoted as $l_1^{i}$, $l_2^{i}$,...,$l_{m_i}^{i}$) in the $i$-th MCG, where $1 \leq i \leq \Omega$. In the $i$-th MCG, a multivariate $m_i$-dimensional log-concave Cumulative Density Function $CDF_i(x_1,x_2,...,x_{m_i})$ is given to allocate cost $x_1$, $x_2$,..., $x_{m_i}$ for links $l_1^{i}$,  $l_2^{i}$,..., $l_{m_i}^{i}$, respectively. 

Therefore, if the possible cost of link $l$ ranges from $0$ to $w_{l}^{\max}$ ($0 <w_{l}^{\max}$), then the probability of allocating a cost value out of this range is $0$. Hence, we have $CDF_{i}(w_{l}^{\max})=1$ for a single element MCG $i$, and $CDF_{j}(w_{l^j_1}^{\max},w_{l^j_2}^{\max},...,w_{l^j_{m_j}}^{\max})=1$ for a multi-element MCG $j$. 

\begin{figure}[tbh]
\centering
\includegraphics[trim = 28mm 93mm 0mm 93mm,clip,width=0.49 \textwidth]{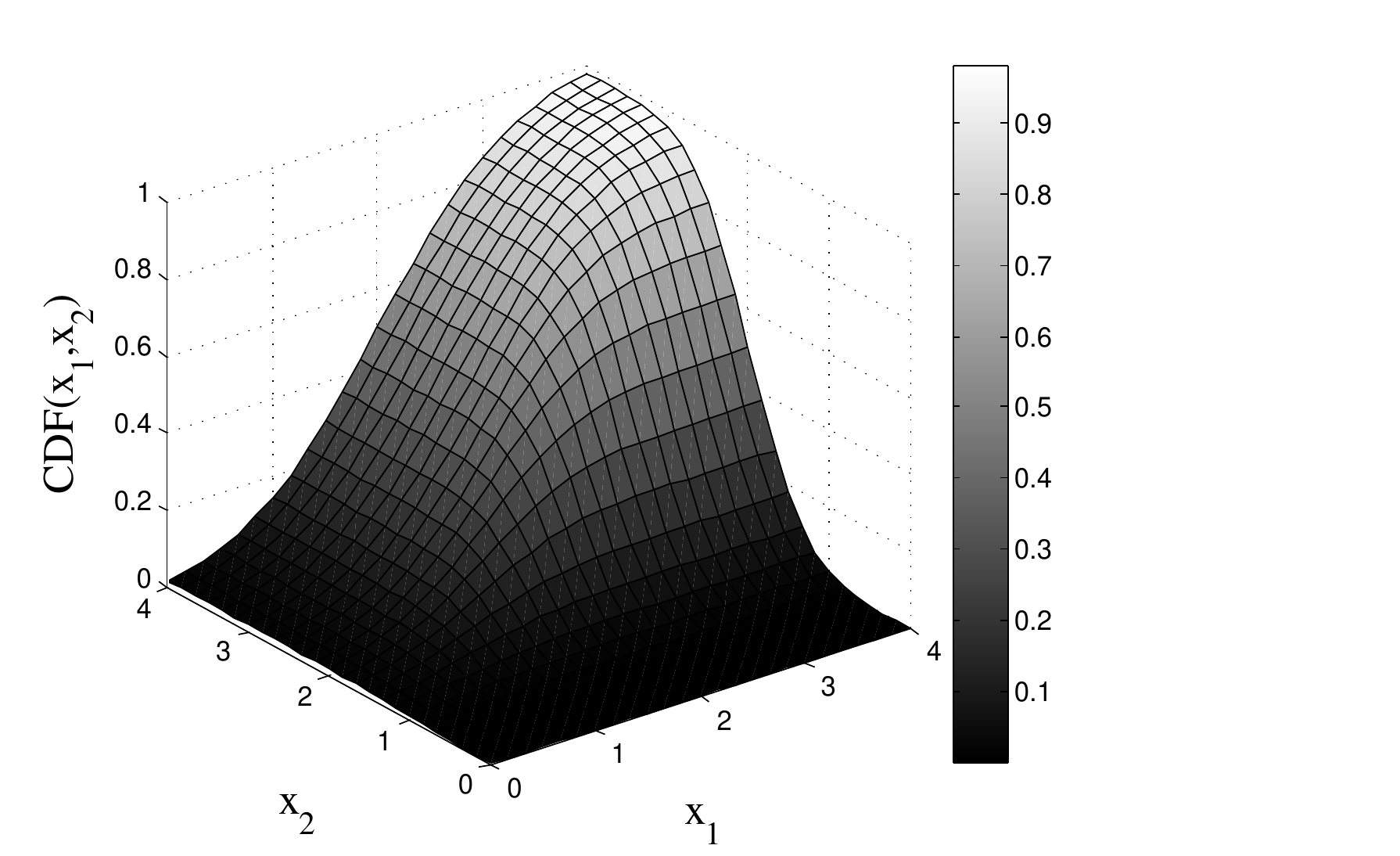}
\caption{A 2-dimensional multivariate Normal distribution.}%
\label{Fig:CvxCDF}%
\end{figure}

For example, Fig.~\ref{Fig:CvxCDF} shows a 2-dimensional multivariate Normal distribution, where both variables are in the range $[0,4]$ with mean $2$ and covariance matrix $\left[
  \begin{array}{cc}
    0.9 & 0.4  \\
    0.4 & 0.3  \\
  \end{array}
\right]$. Similarly to Eq. (\ref{Eq:PbCorreCoe}), the correlation matrix (composed of linear correlation coefficients) can be derived from the covariance matrix and the variables' standard variances in the multivariate Normal distribution. However, we do not explicitly use the \emph{linear} correlation coefficient in the stochastic correlated model, since we will later prove that via the log-concave property of this model, the shortest path problem can be solved by convex optimization.  

\section{Shortest Paths in Correlated Networks} \label{Sec:CorreRouting}
\subsection{Shortest Path under the Deterministic Correlated Model}
\begin{definition}
Given is a directed network $G(\mathcal{N},\mathcal{L})$, and each link $l \in \mathcal{L}$ has a cost $w(l)$ following the deterministic correlated model. The Shortest Path under the Deterministic Correlated Model (SPDCM) problem is to find a path from a source $s$ to a destination $t$ with minimum cost. 
\end{definition} 

In conventional deterministic networks, a subpath of a shortest path is also the shortest. We refer to this property as the dominance of the subpath. However, this is not the case in networks with deterministic correlated link weights, which means a dominated path may also lead to an optimal solution. For instance, in the example of Fig. \ref{Fig: Nondom}, we can see that although subpath $s$-$b$ has a smaller cost than subpath $s$-$a$-$b$, path $s$-$a$-$b$-$t$ (instead of path $s$-$b$-$t$) has minimum cost. In the following, we will study the complexity of the SPDCM problem.    

\begin{theorem} \label{Theo:SPDCM_NP}
The SPDCM problem is NP-hard.
\end{theorem}

\begin{IEEEproof}
When the correlation coefficient is equal to $1$, the SPDCM problem can be solved in polynomial time by a conventional shortest path algorithm. We therefore prove in the following that the SPDCM problem is NP-hard for ``increasing correlation''$>1$ as well as ``decreasing correlation''$<1$.

\textbf{Increasing Correlation}:

When the correlation coefficient is greater than $1$, we make a reduction to the forbidden pairs shortest path problem, which is known to be NP-hard~\cite{Gabow1976}. In a given network and for a given set of node pairs $\zeta$, the forbidden pairs shortest path problem looks for the shortest path between $s$ and $t$ such that at most one node from each pair in the set $\zeta$ lies on this path.
Let us consider a network with deterministic correlated link weights. When two nodes $i$ and $j$ form a forbidden pair, their costs are correlated such that $w(i,.) \oplus w(j,.)=\infty$, where $(i,.)$ and $(j,.)$ represent any link that contains an end node of $i$ and $j$, respectively. In all the other cases, the link costs are uncorrelated and finite. Since $w(i,.) \oplus w(j,.)=\infty$, if the two forbidden nodes appear in the same path then the cost of this path will be $\infty$, so it will never lead to the shortest path. Now, the SPDCM problem is equivalent to the forbidden pairs shortest path problem. 

\textbf{Decreasing Correlation}:

When the correlation coefficient is less than $1$, we make a reduction to the Minimum Color Single-Path (MCSiP) problem, which is NP-hard \cite{Yuan05}. Given a network $G(\mathcal{N},\mathcal{L})$, and given the set
of colors $C=\{c_1,c_2,...,c_g\}$ where $g$ is the total number of colors, and given the color set ${\{c_{l}\}}$ associated to each link $l\in\mathcal{L}$, the Minimum Color
Single-Path (MCSiP) problem is to find one path from source node $s$ to destination node $t$ such that it uses the least amount of colors.

Assume each color $c_i$ is associated with cost $1$, where $1 \leq i \leq g$. We further assume that $w(l_1)\oplus w(l_2) \oplus \cdot \cdot \cdot \oplus w(l_m)=q$, where $q$ is the total number of distinct colors belonging to these $m$ links. Therefore, the SPDCM problem is equivalent to the MCSiP problem. 
\end{IEEEproof}

\begin{theorem} \label{Theo:SPNoAppro}
The SPDCM problem cannot be approximated to arbitrary degree in polynomial time, unless P=NP.
\end{theorem}

\begin{IEEEproof}
We provide a proof by contradiction.

\textbf{Increasing Correlation:}

Assume a polynomial-time approximation algorithm exists that can find a path with a cost at most $\alpha \cdot opt$, where $\alpha>1$ is an approximation ratio. For a pair of forbidden nodes $i$ and $j$, we further assume $w(i,.) \oplus w(j,.)>\alpha * opt$. Therefore, if an approximation algorithm can find a path $\psi$ with cost at most $\alpha * opt$ from $s$ to $t$, then $i$ and $j$ cannot be simultaneously traversed by this path $\psi$, which means that the forbidden pairs shortest path problem can be solved in polynomial time, which results in a contradiction.

\textbf{Decreasing Correlation:}

We first introduce the Disjoint Connecting Paths problem \cite{Garey79}. Given a directed network $G(\mathcal{N},\mathcal{L)}$, a collection of disjoint node pairs ($s_{1}$, $t_{1}$), ($s_{2}$, $t_{2}$), ..., ($s_{z}$, $t_{z}$), does $G$ contain $z$
mutually link-disjoint paths, one connecting $s_{i}$ and $t_{i}$ for each $i$,
$1\leq i\leq z$. This problem is NP-hard when $z \geq 2$. Assume a polynomial-time approximation algorithm exists that can find a path with a cost at most $\alpha \cdot opt$, where $\alpha>1$ is an approximation ratio. Assuming all the links in the network have weight $1$, and link $(u,v)$ and any $m>0$ links in $ \mathcal{L} \backslash \{(u,v)\}$ are correlated, with a total cost of $\frac{1}{\beta} \cdot m$. Moreover, any two or more links in $\mathcal{L} \backslash \{ (u,v) \}$ are assumed to be uncorrelated/independent. 

According to this assumption, the minimum value of a shortest path is $1$ if link $(u,v)$ is not traversed, i.e., it traverses only one link from $s$ to $t$. However, the optimal solution which traverses link $(u,v)$ has a total cost of $opt=\frac{1}{\beta} \cdot c$, where $c$ is the sum of minimum hops from $s$ to $u$ and from $v$ to $t$. For any given $\alpha$, let $\frac{\beta}{c}>\alpha$, then $1>\alpha \cdot \frac{1}{\beta} \cdot c$, which means $1>\alpha \cdot opt$. To find a path with cost at most $\alpha \cdot opt$, the polynomial-time algorithm must find a path which traverses link $(u,v)$. In that case the algorithm can, in polynomial time, find two link-disjoint paths from $s$ to $u$, and from $v$ to $t$, which results in a contradiction.
\end{IEEEproof}

Next, we study the performance of a conventional shortest path algorithm running on a graph where each link has an ``uncorrelated'' weight value.
\begin{lemma} \label{Theo:InRouting}
When all the correlation coefficients are greater than $1$, a conventional shortest path $\psi$ has a total cost at most $\frac{\rho_{\max}}{\rho_{opt}} \cdot opt$, where $\rho_{\max}$ and $\rho_{opt}$ are the largest correlation coefficient and the optimal solution's correlation coefficient, respectively, and $opt$ is the cost of the optimal solution.
\end{lemma}
\begin{IEEEproof}
Let $U(\psi)=\sum_{l \in \psi} w(l)$ and let $C(\psi)=\rho_u \cdot U(\psi) =\rho_{u} \cdot \sum_{l \in \psi} w(l)$ reflect the total joint cost of path $\psi$ considering their correlation, where $\rho_u$ indicates the correlation coefficient of path $\psi$. On one hand, a conventional shortest path $\psi$ should satisfy $U(\psi) \leq \frac{opt}{\rho_{opt}}$. On the other hand, $C(\psi) \leq \rho_{\max} \cdot U(\psi)$ considering $\rho_{max}$ is the largest correlation coefficient. Hence, $C(\psi)\leq \rho_{\max} \cdot U(\psi)\leq \frac{\rho_{\max}}{\rho_{opt}} \cdot opt$. 


\end{IEEEproof}

\begin{lemma} \label{Theo:DeRouting}
When all the correlation coefficients are less than $1$, a conventional shortest path $\psi$ has a cost at most $\frac{1}{\rho_{\min}} \cdot opt$, where $\rho_{\min}$ is the smallest correlation coefficient among all the correlation coefficients.
\end{lemma}
\begin{IEEEproof}
Let $V(\psi)=\sum_{l \in \psi} w(l)$ and let $C(\psi)=\rho_{u} \cdot \sum_{l \in \psi} w(l)$ reflect the total joint cost of path $\psi$ considering their correlation. Since all the correlations are decreasing ($\rho<1$), we have $C(\psi) \leq V(\psi)$. On the other hand, $\rho_{\min} \cdot V(\psi) \leq opt$ considering that $\rho_{\min}$ is the smallest correlation coefficient. Hence, $C(\psi) \leq V(\psi) \leq \frac{1}{\rho_{\min}} \cdot opt$.
\end{IEEEproof}

Via Lemmas \ref{Theo:InRouting} and \ref{Theo:DeRouting}, we obtain Theorem \ref{Theo:ApproRouting}. 
\begin{theorem} \label{Theo:ApproRouting}
In a network with links following the deterministic correlated model, a conventional shortest path can have cost at most $max(\frac{\rho_{\max}}{\rho_{opt}}, \frac{1}{\rho_{\min}}) \cdot opt$.
\end{theorem}

Theorem~\ref{Theo:ApproRouting} reveals that a conventional shortest path may have arbitrary bad performance, since either $\frac{\rho_{\max}}{\rho_{opt}}$ can be infinitely large or $\rho_{min}$ can be infinitely small. 

\subsection{An Exact Algorithm to Solve the SPDCM problem}

To solve the SPDCM problem exactly, we modify Dijkstra's algorithm by letting each node store as many subpaths as possible, which is similar to the exact algorithm for solving the multi-constrained routing problem \cite{SAMCRA}. Since each node can store as many subpaths as possible, its running time is exponential. We start with some notations used in the algorithm: \\

$sus[u][h]$: the parent node of the $h$-th subpath from $s$ to $u$. 

$dist[u][h]$: the cost value of the $h$-th subpath from $s$ to $u$. 

$counter[u]$: the number of stored subpaths at node $u$. 

$u[m]$: the $m$-th subpath from $s$ to $u$. 

$adj(u)$: the set of nodes adjacent to node $u$.   \\

The pseudo-code of the exact algorithm is given in Algorithm \ref{Alg: ExactSPDCM}. 

\begin{algorithm}[!h]
\caption{$SPDCM(G, s, t)$}
\begin{algorithmic}[1] \label{Alg: ExactSPDCM}
\STATE  $Q \leftarrow s$, $P \leftarrow \emptyset$, $dist[s][1] \leftarrow 0$, $dist[i][h] \leftarrow \infty$, $sus[i][h] \leftarrow i$, $counter[s] \leftarrow 1$, $counter[i] \leftarrow 0$, $ \forall i \in \mathcal{N} \backslash \{s\}$.
\STATE \textbf{While $Q \neq \emptyset $} 
\STATE \quad $u[m] \leftarrow$ Extract-min($Q$)
\STATE \quad \textbf{If} ($u==t$) \textbf{do}
\STATE \quad \quad  Insert ($P$, $u[m]$)
\STATE \quad \textbf{Else}
\STATE \quad \quad \textbf{Foreach} $v \in adj(u)$ \textbf{do}
\STATE \quad \quad \quad $counter(v)=counter(v)+1$
\STATE \quad \quad \quad Calculate the total cost of subpath $u[m] \rightarrow v$ and assign it to $dist[v][counter(v)]$
\STATE \quad \quad \quad  $sus[v][counter(v)] \leftarrow u$
\STATE \quad \quad \quad  Insert ($Q,v,counter(v)$)
\STATE return min($P$).
\end{algorithmic}
\end{algorithm}

The time complexity of Algorithm~\ref{Alg: ExactSPDCM} can be computed as follows. Let $k_{\max}$ denote the maximum number of subpaths for each node to store, then in Step $2$, $Q$ contains at most $k_{\max}N$ subpaths.
According to \cite{VanMieghem01}, $k_{\max} \leq \lfloor  e(N-2)! \rfloor$, where $e \approx 2.718$. When using a Fibonacci heap to structure the heap, selecting the minimum cost path has a time complexity of $O(\log(k_{\max}N))$ \cite{Cormen01} in Step $3$. Step $7$-Step $11$ take at most $O(k_{\max})$ time for each link to be iterated and hence result in $O(k_{\max}L)$ time; because for a fixed link, the steps within the inner loop (Steps $8$-$11$) all cost $O(1)$ time. Step $12$ invokes $O(k_{\max})$ time for node $t$ to select the minimum cost path. Hence, the overall time complexity of Algorithm~\ref{Alg: ExactSPDCM} is $O(k_{\max}N \log(k_{\max}N)+k_{\max}L)$.   

\subsection{Shortest Path under the Nodal Deterministic Correlated Model}
\label{Sec:PathNodalCorrelation}

In some real-world networks (e.g., SRLG networks), the links that are spatially (geographically) close to each other are usually correlated, whereas the links that are located far from each other are usually uncorrelated. 
We make an additional assumption, which is that only the links sharing the same node can be correlated. We call this \emph{nodal} correlation. 

Although the SPDCM problem is NP-hard, we will show that, by transforming the original graph to an auxiliary graph, the Shortest Path under the Nodal Deterministic Correlated Model (SPNDCM) problem is solvable in polynomial time. For any node $a$, there are generally two cases of nodal correlation, namely (1) links in the form of $(a,b)$ and $(a,c)$, and (2) links in the form of $(a,b)$ and $(b,c)$ are correlated. When $(a,b)$ and $(a,c)$ are correlated, a simple path cannot traverse both of them, since looping is not allowed. In this sense, any simple path only traverses at most one of them, which means that the links' correlation will not affect the cost calculation of any simple path. Therefore, we only need to consider the case when $(a,b)$ and $(b,c)$ are correlated. We first define that if $(a,b)$ and $(b,c)$ are correlated, then $a$ and $b$ are called \emph{correlated nodes}, which is represented by $C_n$, else they are \emph{uncorrelated nodes}, which is denoted by $U_n$. Subsequently, based on the original graph $G(\mathcal{N},\mathcal{L})$, the auxiliary graph $G^{A}(\mathcal{N}^{A},\mathcal{L}^{A})$ can be constructed as follows:
\begin{enumerate}

\item For any two links $(u,v) \in \mathcal{L}$ and $(v,y) \in \mathcal{L}$ that are correlated in $G$, create new nodes $u_v$, $v^{uy}$, $v_y$ and $y_y$ in $G^{A}$ if they do not already exist. For node $v \in \mathcal{N}$ in $\mathcal{G}$, in case $v^{uy}$ and $v_v$ need to be created, create $v^{uy}$ only and regard $v^{uy}$ to be the same as $v_v$.  

\item For any node $a \in \mathcal{N}$ and if it is an \emph{uncorrelated node} (in $U_n$), create node $a_a$ in $G^{A}$.

\item For any two correlated links $(u,v)$ and $(v,y)$ in $G$, create links $(u_v,v^{uy})$, $(v^{uy},v_y)$ and $(v_y,y_y)$ in $\mathcal{G}^{A}$. Assign to the links $(u_v,v^{uy})$ and $(v_y,y_y)$ the weights of $w(u,v)$ and $w(v,y)$, respectively, and the link $(v^{uy},v_y)$ with weight $(\rho_{(u,v)(v,y)}-1) \cdot (w(u,v)+w(v,y))$, where $\rho_{(u,v)(v,y)}$ is the correlation coefficient of links $(u,v)$ and $(v,y)$.

\item For each link $(a,b) \in \mathcal{L}$ such that both node $a$ and node $b$ are not \emph{correlated nodes}, create the link $(a_a,b_b)$ also in $G^{A}$ with the link weight of $w(a,b)$.

\item For each link $(a,b) \in \mathcal{L}$ such that $a \in U_n$ and $b \in C_n$, draw links $(a_a, b_r)$ in $G^A$, where $r \in \mathcal{N}$ and $b_r \in G^A$.

\item For each link $(a,b) \in \mathcal{L}$ such that $a \in C_n$ and $b \in U_n$, draw links $(a^{rz}, b_b)$ in $G^A$, where $r,z \in \mathcal{N}$ and $a^{rz} \in \mathcal{N}^A$.  

\end{enumerate}

\begin{figure}[tbh]
\centering
\includegraphics[trim = 0mm 0mm 0mm 0mm,clip,width=0.43\textwidth]{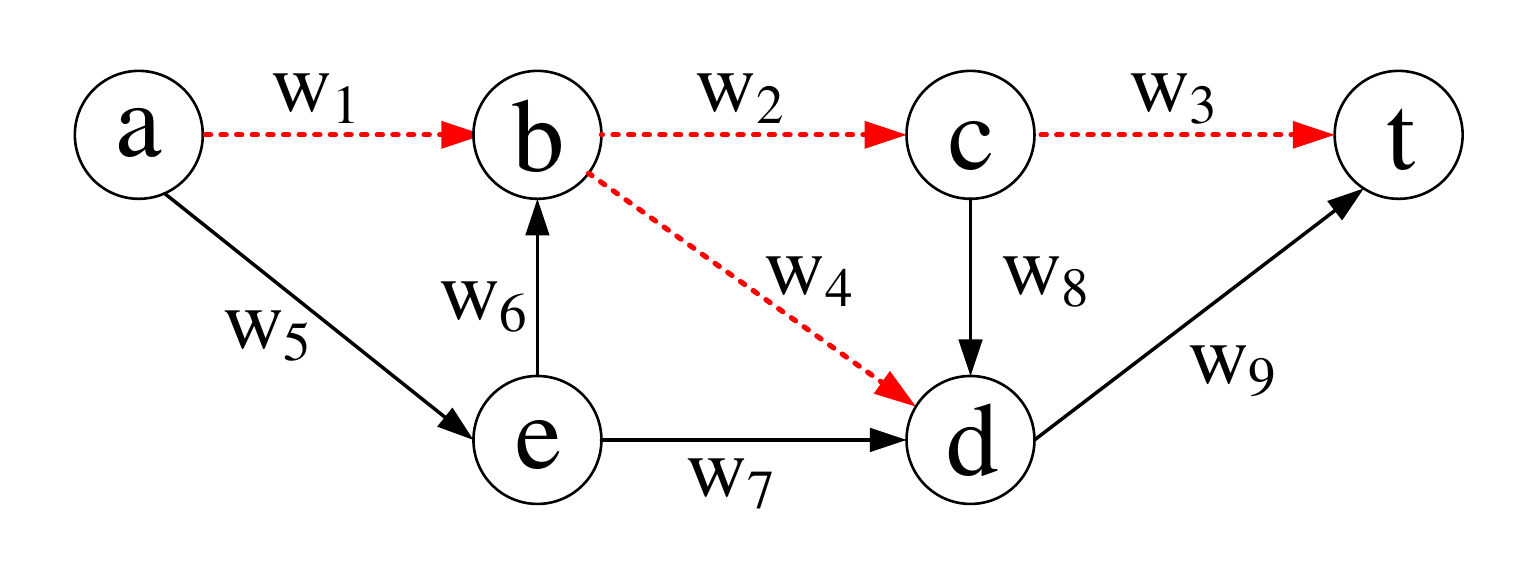}
\caption{An example network with dotted links following the nodal deterministic correlated model.}%
\label{Fig:AuxExp}%
\end{figure}

\begin{figure}[tbh]
\centering
\includegraphics[trim = 0mm 0mm 0mm 0mm,clip,width=0.51\textwidth]{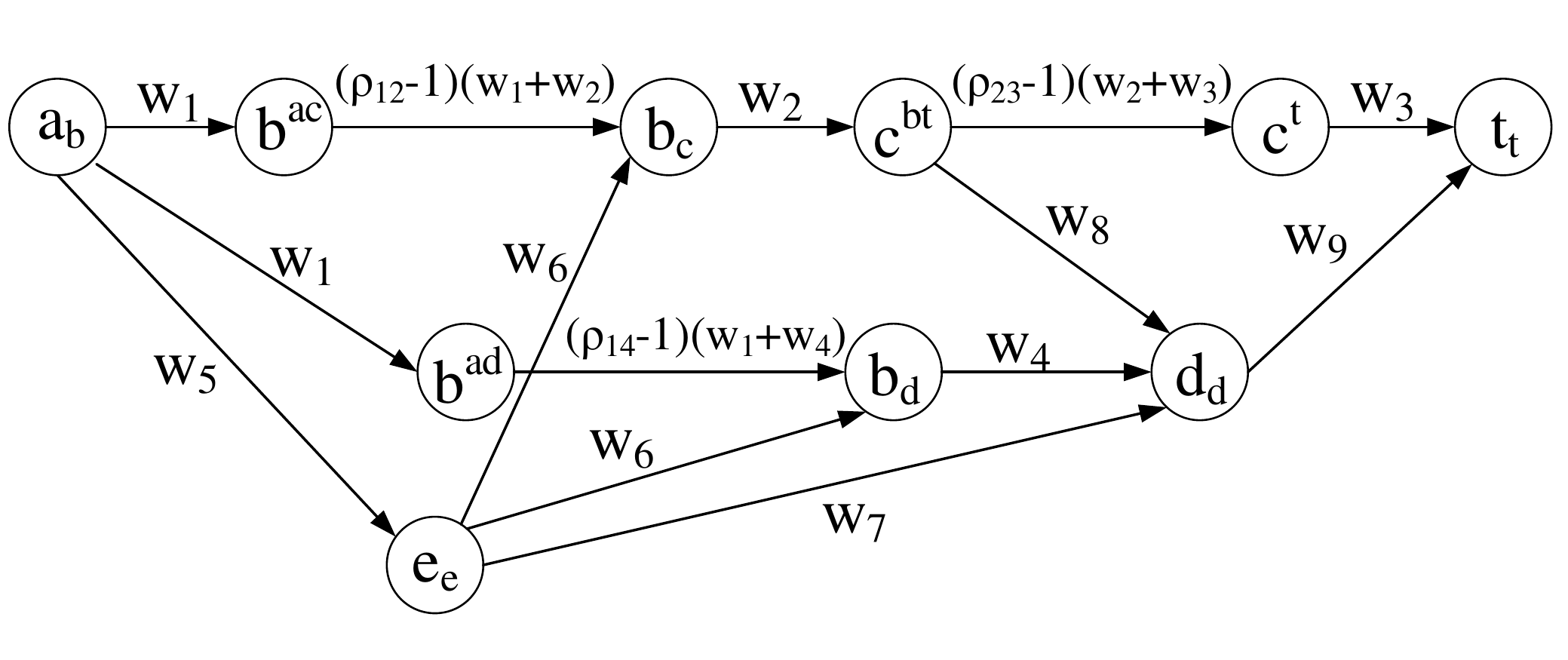}
\caption{Auxiliary graph of Fig.~\ref{Fig:AuxExp} for the SPDCM problem under the nodal deterministic correlated model.}%
\label{Fig:AuxGrap}%
\end{figure}

The idea of the auxiliary graph is that if two links $(u,v)$ and $(v,y)$ are correlated, we create four corresponding nodes $u_v$, $v^{uy}$, $v_y$, $y_y$ and then draw three links $(u_v,v^{uy})$,$(v^{uy},v_y)$ and $(v_y,y_y)$. We use $(u_v,v^{uy})$ and $(v_y,y_y)$ to indicate their uncorrelated values, respectively, if only one of these two links is traversed, and $(v^{uy},v_y)$ to represent the correlated loss (decreasing correlation) or gain (increasing correlation), respectively, if they are traversed simultaneously. For instance, in Fig.~\ref{Fig:AuxExp} where the link weight is labeled above each link, assuming links $(a,b)$, $(b,c)$, $(c,t)$ and $(b,d)$ are nodal correlated, then Fig.~\ref{Fig:AuxGrap} is its corresponding auxiliary graph with the assigned weight shown on each link. In particular, since $(a,b)$, $(b,c)$ and $(a,b)$, $(b,d)$ may have different correlation coefficients, in Fig.~\ref{Fig:AuxGrap} we use $(b^{ac},b_c)$ and $(b^{ad},b_d)$ to represent their correlation value. Meanwhile, when there is a link from an \emph{uncorrelated node} to a \emph{correlated node} in the original graph, e.g., $(e,b)$ in Fig.~\ref{Fig:AuxExp}, we draw links $(e_e,b_c)$ and $(e_e,b_d )$ (Step 5). When there is a link from a \emph{correlated node} to an \emph{uncorrelated node} in the original graph, e.g., $(c,d)$ in Fig.~\ref{Fig:AuxExp}, we draw link $(c^{bt},d_d)$ (Step 6). Considering that there are at most $N(N-1)$ nodal links in a graph, the original graph can be transferred to the auxiliary graph in polynomial time. 

Consequently, running a shortest path algorithm on the auxiliary graph can return a minimum cost path under the nodal deterministic correlated model. Our auxiliary graph can deal with both decreasing and increasing correlation cases. Considering that $(\rho_{(u,v)(v,y)}-1) \cdot (w(u,v)+w(v,y))<0$ in the auxiliary graph under the decreasing correlation case, and Dijkstra's algorithm cannot handle negative link weights, we could for instance run Bellman-Ford's algorithm on the auxiliary graph. No negative loops will exist in the auxiliary graph, since if a path traverses the negative weight link (say $(v^{uy},v_y)$), it will also traverse the links in the form of $(u_v,v^{uy})$ and $(v_y,y_y)$, whose total cost is always positive.  


\subsection{Shortest Path under the Stochastic Correlated Model} \label{Sec:SPSCM}

\begin{definition}
The Shortest Path under the Stochastic Correlated Model (SPSCM) problem: In a given directed graph $\mathcal{G}(\mathcal{N},\mathcal{L})$ where the link costs follow the stochastic correlated model, it is assumed that there are in total $\Omega$ Maximum Correlated Groups (MCGs). The SPSCM problem is to find a path from source $s$ to destination $t$ such that its total cost is minimized and the probability to realize this value is no less than $P_s$. 
\end{definition}


We present a convex optimization formulation to solve the SPSCM problem. Convex optimization problems can usually be solved quickly and accurately with convex optimization solvers \cite{boyd04}. Let us first introduce how to develop a Linear Programming (LP) formulation to solve the shortest path problem in deterministic networks:

\textbf{Objective:}
\begin{equation}
\min \sum_{(u,v) \in \mathcal{L}}w(u,v) \cdot y_{uv}
\end{equation}

\textbf{Constraints:}

\begin{equation}
y_{uv} \geq 0
\end{equation}

\begin{equation}
\sum_{v \in\mathcal{N}} y_{uv} - \sum_{v \in\mathcal{N}}
y_{uv}=\left\{
\begin{array}
[c]{c}%
1\\
-1\\
0
\end{array}
\right.  \text{ }%
\begin{array}
[c]{l}%
u=s\\
u=t\\
otherwise
\end{array}
\label{Eq:SP}%
\end{equation}
where $y_{uv}$ indicates whether link $(u,v)$ is part of the shortest path. When $y_{uv}=1$, it indicates that link $(u,v)$ appears on the path, else $y_{uv}=0$. The objective is to minimize the total cost value of the path. Constraint Eq.~(\ref{Eq:SP}) accounts for that except for $s$ and $t$, the number of incoming and outgoing links that are part of the path must be the same. For the source node $s$, the number of its outgoing links should be $1$, and for the destination node $t$ the number of its incoming links should be $1$. The dual of the above Linear Program (LP) can be expressed as follows:  	

\textbf{Objective:}
\begin{equation}
\max \text{ } d_t
\end{equation}

\textbf{Constraints:}

\begin{equation}
d_s=0
\end{equation}

\begin{equation}
d_v-d_u \leq w(u,v) \text{ } \text{ } \forall (u,v) \in \mathcal{L}
\end{equation}
where $d_u$ is a value between $0$ and $1$. Similarly, the SPSCM problem can be solved by the following convex formulation:

\textbf{Objective:}
\begin{equation}
\max \text{ } d_t
\end{equation}

\textbf{Constraints:}

\begin{equation}
d_s=0
\end{equation}

\begin{equation}
d_v-d_u \leq x(u,v) \text{ } \text{ } \forall (u,v) \in \mathcal{L}
\end{equation}

\begin{equation}
\label{Eq:RoutingProb}
\sum_{i \in \Omega} -\log\left( CDF_i(x(l^i_1),x(l^i_2),...,x(l^i_{m_i}))\right) \leq - \log (P_s)
\end{equation}
where the variables $x(u,v),x(l^i_1),x(l^i_2),...,x(l^i_{m_i})$ indicate the allocated possible cost of links $(u,v)$, $l^i_1$, $l^i_2$,..., $l^i_{m_i}$, respectively. Constraint~(\ref{Eq:RoutingProb}) ensures that the total probability of realizing the total cost is no more than $P_s$. In Eq.~(\ref{Eq:RoutingProb}), for each MCG we apply the multi-dimensional CDF functions to calculate the probability of realizing a cost. Since the multi-dimensional CDF function is log-concave, $-\log\left( CDF_i(x(l^i_1),x(l^i_2),...,x(l^i_{m_i}))\right)$ is convex, and by summing all the MCGs' CDFs together, it remains convex, which indicates that Eq.~(\ref{Eq:RoutingProb}) is convex. The other constraints are also convex, which proves that the above formulation is a convex optimization formulation. 

\subsection{Widest Path under the Deterministic Correlated Model}

The Widest Path in Deterministic Networks (WPDN) problem is to find a path from $s$ to $t$ such that the minimum link weight among all its traversed links is maximized. This problem appears with bottleneck metrics, such as bandwidth. The WPDN problem is solvable in polynomial time: First, we order all the link weights in the network in increasing order. After that, each round we prune lowest-weight links in the graph and run a Depth First Search (DFS) or a Breadth First Search (BFS) algorithm to find a path from $s$ to $t$. The algorithm will end if there is no path anymore from $s$ to $t$ and return the pruned weight value of the previous round. 

In the Widest Path under the Deterministic Correlated Model (WPDCM) problem, if $m>1$ correlated links in a path have a joint weight value $W$, then for each link the maximum average/amortized weight is $\frac{W}{m}$. For instance, if a path traverses three correlated links with joint weight value of $15$ and passes another uncorrelated link with weight of $6$, then this path has a ``width'' value of $5$. The reason is that the maximum (average/amortized) weight for each of these three correlated links is $15/3=5$, and this value is less than for another uncorrelated link ($6$).  

However, the WPDCM problem is still NP-hard and cannot be approximated to arbitrary degree. The proof follows analogously from Theorems \ref{Theo:SPDCM_NP} and \ref{Theo:SPNoAppro}.

\section{Minimum Cuts in Correlated Networks} \label{Sec:CorreCut}
\subsection{Min-Cut under the Deterministic Correlated Model} 
\begin{definition}
The Min-Cut under the Deterministic Correlated Model (MCDCM) problem: Given is a network $G(\mathcal{N},\mathcal{L})$, and each link $l \in \mathcal{L}$ is associated with a cost $w(l)$. It is assumed that two or more link costs are correlated under the deterministic correlated model. Given a source $s$ and a target $t$, find a cut $\mathcal{C}$ that partitions $G$ into two disjoint subsets $X$ ($X \in \mathcal{N}$) and $\mathcal{N}-X$ such that $s$ and $t$ are in different subsets and the cost of the cut $C$ is minimized.
\end{definition} 
\begin{figure}[tbh]
\centering
\includegraphics[trim = 0mm 0mm 0mm 0mm,clip,width=0.23\textwidth]{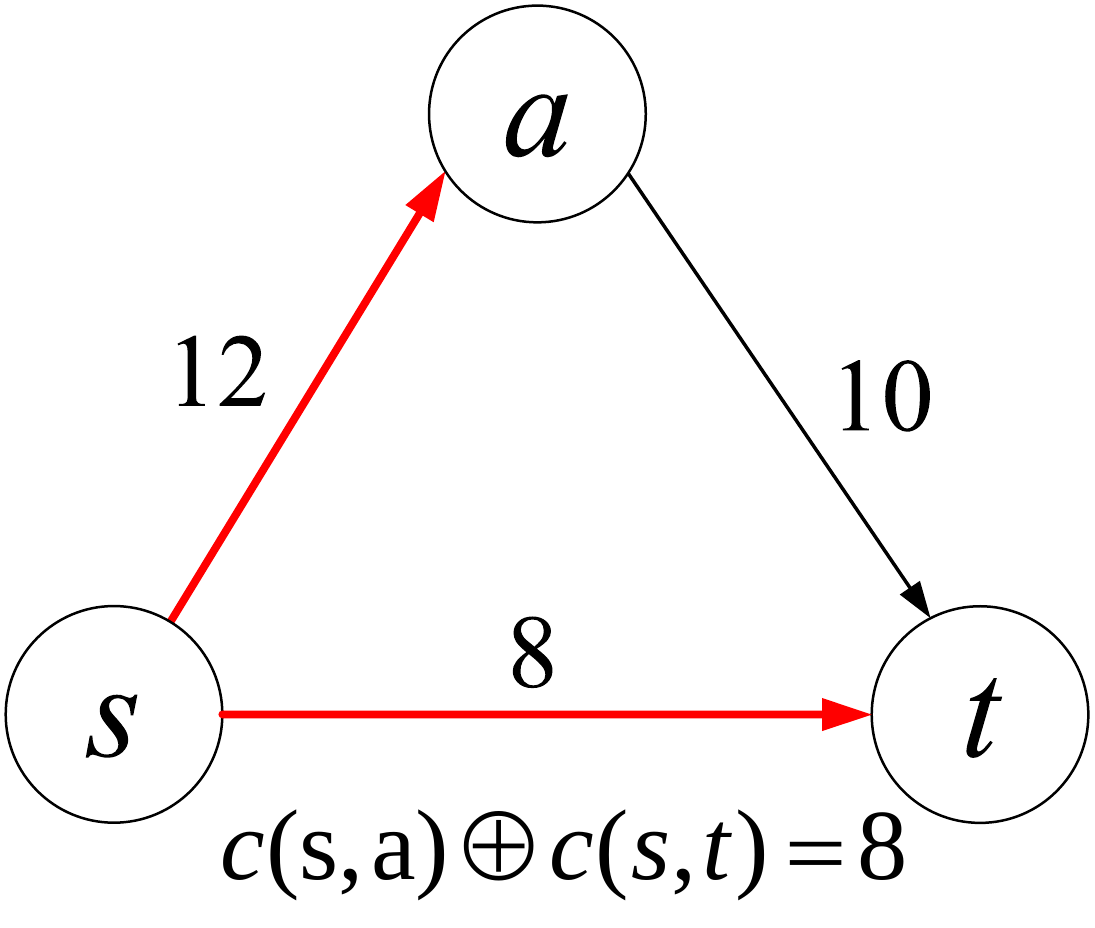}
\caption{An example to illustrate that the maximum flow is not equal to the min-cut in correlated networks.}%
\label{Fig: MaxFlowMinCut}%
\end{figure}

The min-cut value is not equal to the maximum flow value under the deterministic correlated model. For example, in Fig.~\ref{Fig: MaxFlowMinCut} assume links $(s,a)$ and $(s,t)$ are correlated with joint cost of $8$. In this example, the maximum flow from $s$ to $t$ is $s-a-t$ with value $10$, while the min-cut is composed of links $(s,a)$ and $(s,t)$ and has a cost of $8$.
  
\begin{figure}[tbh]
\centering
\includegraphics[trim = 0mm 0mm 0mm 0mm,clip,width=0.21\textwidth]{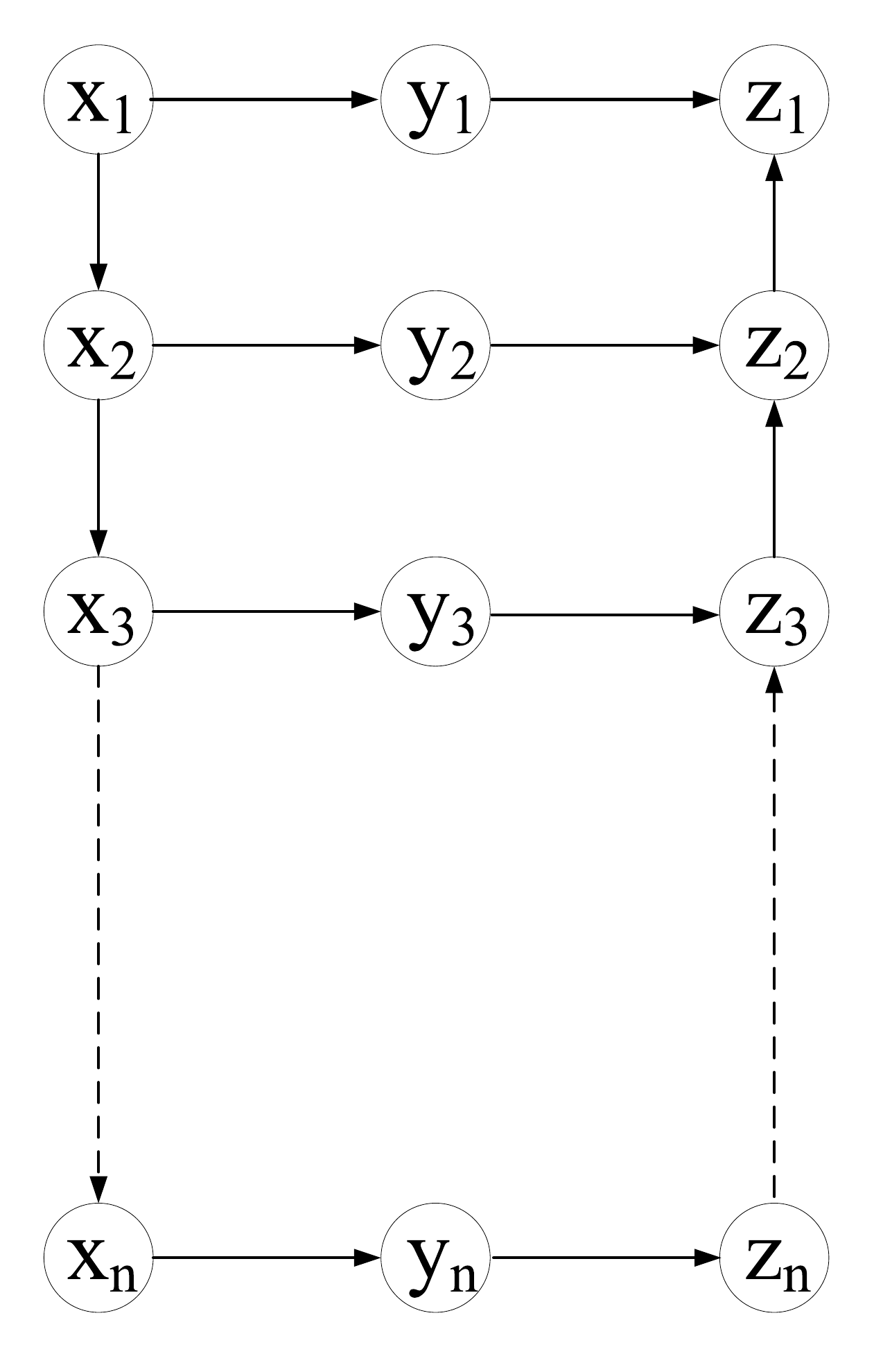}
\caption{NP-hardness of the MCDCM problem.}%
\label{Fig: NPHardCNC}%
\end{figure}

\begin{figure}[tbh]
\centering
\includegraphics[trim = 0mm 0mm 0mm 0mm,clip,width=0.43 \textwidth]
{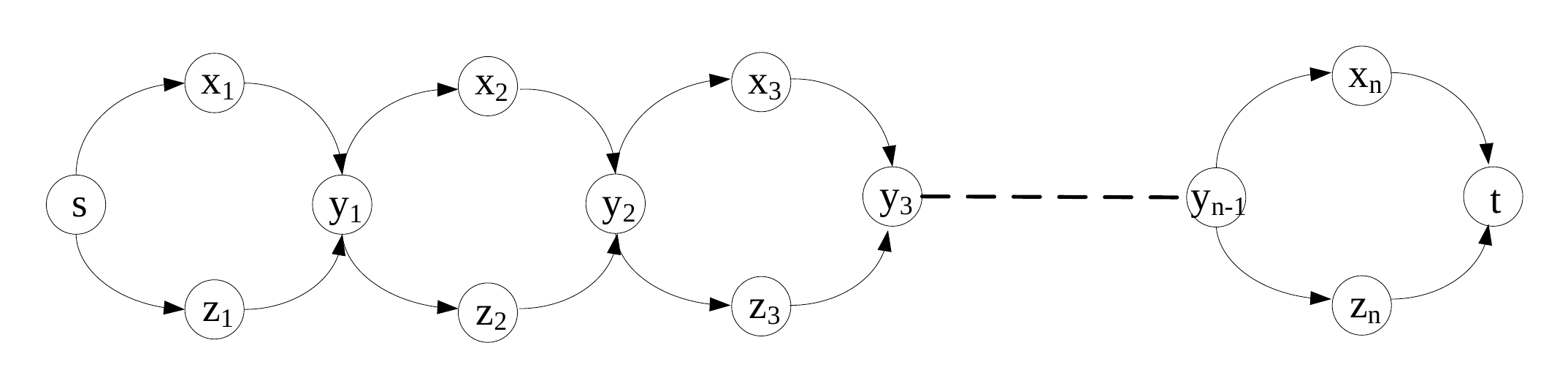}
\caption{A reduction of the MCDCM problem to the SPDCM problem.}%
\label{Fig: NPHardCorrelated}%
\end{figure}

\begin{theorem} \label{Theo: MCDCM_NPhard}
The MCDCM problem is NP-hard.
\end{theorem}

\begin{IEEEproof}
In Fig.~\ref{Fig: NPHardCNC}, we assume that the links in the form of $(x_i,x_{i+1})$ and $(z_{i+1},z_{i})$ have infinite uncorrelated cost and the link costs of $(x_i,y_i)$ and $(y_i,z_i)$ follow the deterministic correlated model, where $1 \leq i \leq n$. We want to find a min-cut to separate $x_1$ and $z_1$. Based on Fig.~\ref{Fig: NPHardCNC}, we first derive Fig. \ref{Fig: NPHardCorrelated} with the same nodes except that we add one more node $s$. We set $s=y_0$, and $t=y_{n}$. The link weight in Fig. \ref{Fig: NPHardCorrelated} is set as follows: $(y_{i-1}, x_{i})$ and $(y_{i-1}, z_{i})$ have $0$ uncorrelated cost, while $(x_{i}, y_{i})$ and $(z_{i}, y_{i})$ have the same (correlated) costs with $(x_i, y_i)$ and $(y_i,z_i)$ in Fig. \ref{Fig: NPHardCNC}, respectively, where $1 \leq i \leq n$. In Fig. \ref{Fig: NPHardCorrelated}, we want to solve the SPDCM problem from the source $s$ to the destination $t$. 

Since we want to find a min-cut that separates $x_1$ and $z_1$, any cut in the form of $(x_i,y_i)$ and $(y_i,z_i)$, where $1 \leq i \leq n$, is not the optimal solution. The reason is that this kind of cut only separates $y_i$ and other nodes, but not $x_1$ and $z_1$. Moreover, considering the links in the form of $(x_j,x_{j+1})$ or $(y_j,y_{j+1})$ have infinite costs, they cannot be in the optimal solution. Based on above analysis, any feasible cut $C$ should contain one link of either $(x_i,y_i)$ or $(y_i,z_i)$, for all $1 \leq i \leq n$. We prove in the following that the MCDCM problem in Fig. \ref{Fig: NPHardCNC} can be reduced to the SPDCM problem in Fig. \ref{Fig: NPHardCorrelated} in polynomial time.

The SPDCM problem to the MCDCM problem: Considering an optimal solution of the SPDCM problem, and denote $R_{SPDCM}$ as the set of links in the optimal solution of the SPDCM problem. Because $R_{SPDCM}$ has minimum cost, let $C_{MCDCM}=R_{SPDCM} \backslash \{(y_i,x_{i+1}), (y_i,z_{i+1})\}$ and then $ \forall (z_i,y_i) \in C_{MCDCM} $, change it to $(y_i,z_i)$ in $C_{MCDCM}$. Since the links $(y_i,x_{i+1})$ and $(y_i,z_{i+1})$ have $0$ cost, $C_{MCDCM}$ also has minimum cost (the same with $R_{SPDCM}$). Therefore solving the SPDCM problem yields a solution to the MCDCM problem. 

The MCDCM problem to the SPDCM problem: An optimal solution of the MCDCM problem should be composed of either $(x_i, y_i)$ or $(y_i,z_i)$, where $1 \leq i \leq n$. Denote $C_{MCDCM}$ as the set of links in the optimal solution of the MCDCM problem. Let $R_{SPDCM}=C_{MCDCM} $ and then $ \forall (y_i,z_i) \in R_{SPDCM} $, change it to $(z_i,y_i)$ in $R_{SPDCM}$. Because $C_{MCDCM}$ has minimum cost value and the links in the form of $(y_{i-1},x_{i})$ or $(y_{i-1},z_{i})$ have $0$ cost, $R_{SPDCM}$ together with $(y_{i-1},x_{i})$ if $(x_i,y_i) \in R_{SPDCM} $ or $(y_{i-1},z_{i})$ if $(z_i,y_i) \in R_{SPDCM} $, can form a path from $s$ to $t$ with minimum cost. Hence, a solution to the MCDCM problem can also solve the SPDCM problem. 
\end{IEEEproof}

\begin{theorem}
The MCDCM problem cannot be approximated to arbitrary degree in polynomial time, unless P=NP.
\end{theorem}
\begin{IEEEproof}
The proof follows from the fact that the SPDCM problem cannot be approximated to arbitrary degree in polynomial time according to Theorem \ref{Theo:SPNoAppro}.
\end{IEEEproof}

\begin{theorem}
By assigning each link $l$ with the cost $w(l)$, running a conventional min-cut algorithm will return a cut with total cost at most $\max(\frac{\rho_{\max}}{\rho_{opt}}, \frac{1}{\rho_{\min}}) \cdot opt$.
\end{theorem}
\begin{IEEEproof}
The proof follows analogously from Theorem \ref{Theo:ApproRouting}. 
\end{IEEEproof}

Since the MCDCM problem is NP-hard and even does not admit a polynomial-time approximation algorithm, we suggest a brute-force approach to solve it. The idea is that we start with two sets $A$ and $B$, with $s$ in $A$ and $t$ in $B$. Then we have $N-2$ nodes left, and there are ${N-2 \choose 0}+{N-2 \choose 1}+ \cdot \cdot \cdot +{N-2 \choose N-2}=O(2^{N})$  combinations to assign these $N-2$ nodes to sets $A$ and $B$. Each combination assignment corresponds to a cut to separate $A$ and $B$, and the one with minimum cost is returned as the optimal solution. 
\subsection{Min-Cut under the SRLG-like Correlated Model}
In Section~\ref{Sec:CorreModels}, we introduced and formulated the joint failure calculation in SRLG networks, which follows the decreasing correlated model. We define the SRLG-like correlated model as follows:
\begin{definition}
The SRLG-like correlated model: Suppose $l_1$, $l_2$,...,$l_m$ ($1<m \leq L$) form a Correlated Group (CG), then $w(l_1) \oplus w(l_2) \cdot \cdot \cdot \oplus w(l_j)$ is greater than the sum of at most $j-1$ link costs, but smaller than $w(l_1)+ w(l_2) \cdot \cdot \cdot + c(l_j)$, where $1< j \leq m$. \\
\end{definition}

The Shortest Path under the SRLG-like model (SP-SRLG) problem is NP-hard, since it is a general case of the MCSiP problem introduced in Section~\ref{Sec:CorreRouting}, which is NP-hard \cite{Yuan05}. Also the Min-Cut under the SRLG-like correlated model (MC-SRLG) problem is NP-hard. Similar to the proof that the MCDCM problem is NP-hard in Section~\ref{Sec:CorreCut}, the MC-SRLG problem in the form of Fig.~\ref{Fig: NPHardCNC} can be reduced to the NP-hard SP-SRLG problem. 

\begin{figure}[tbh]
\centering
\includegraphics[trim = 0mm 10mm 0mm 0mm,clip,width=0.25 \textwidth]{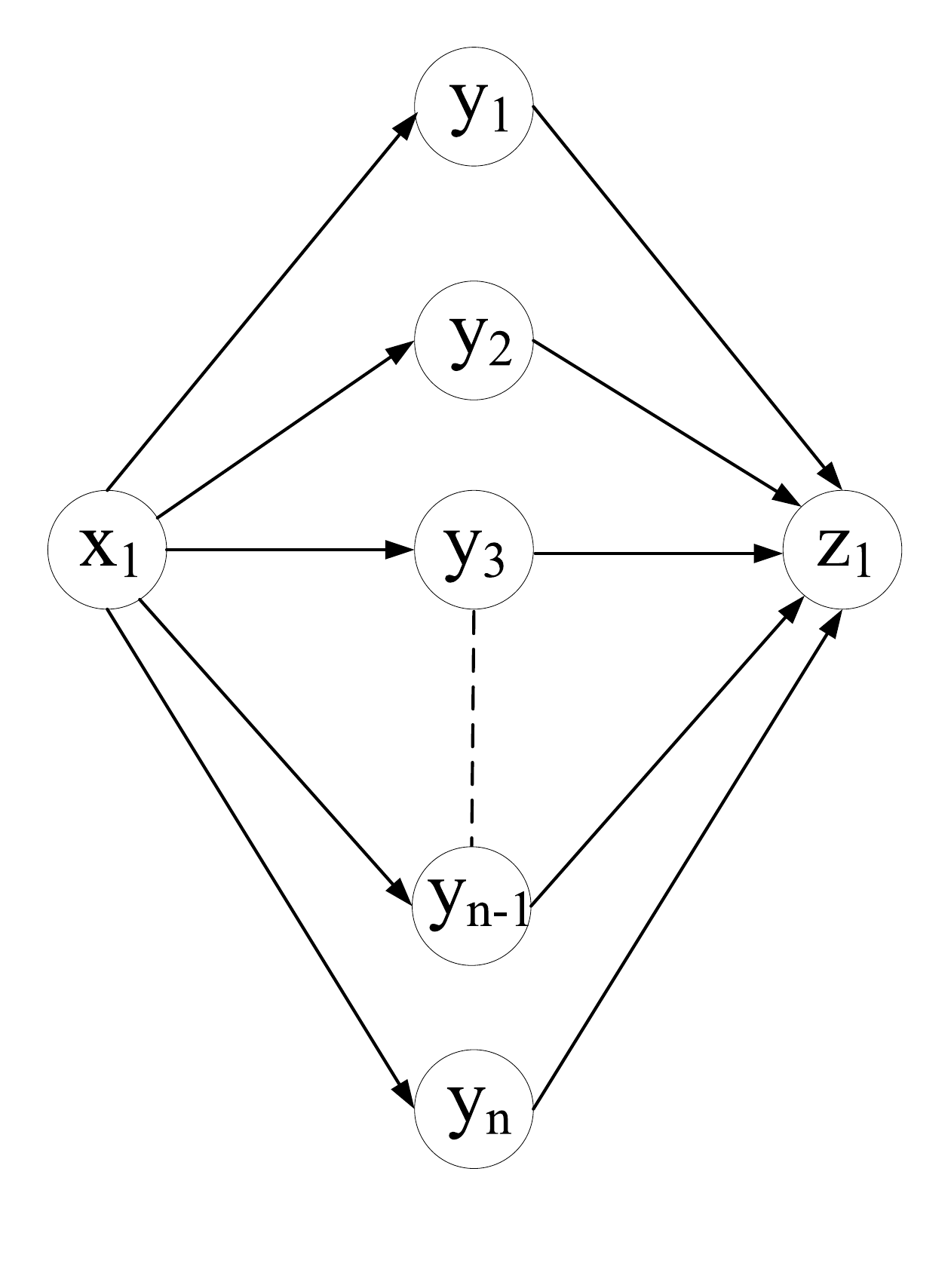}
\caption{NP-hardness of the MC-NSRLG problem.}%
\label{Fig: NPHardPairCNC}%
\end{figure}

In the Nodal SRLG-like correlated model, we assumed that only the links that share the same node follow the SRLG-like correlated model. As Section \ref{Sec:PathNodalCorrelation} shows that the Shortest Path under the Nodal Deterministic Correlation Model problem is solvable in polynomial time, we address the Min-Cut under the Nodal SRLG-like correlated model (MC-NSRLG) problem in the following. In general, the MC-NSRLG problem is still NP-hard. The reason is that for the MC-NSRLG problem in Fig.~\ref{Fig: NPHardPairCNC}, where the links in the form of $(x_1,y_i)$ are assumed to be correlated, we could derive a graph like in Fig.~\ref{Fig: NPHardCorrelated} by duplicating $n-1$ more $x$ nodes, $z$ nodes and one $y$ node. We set $s=y_0$ and $t=y_n$. The link weights in Fig.~\ref{Fig: NPHardCorrelated} are set as follows: For any two $(x,y_i)$ and $(x,y_j)$ (or $(y_i,z_1)$ and $(y_j,z_1)$) in Fig.~\ref{Fig: NPHardPairCNC}, $(x_i,y_i)$ and $(x_j,y_j)$ (or $(y_i,z_i)$ and $(y_j,z_j)$) in Fig.~\ref{Fig: NPHardCorrelated} follow the same correlation. This link weight setting also applies to more than two links. The link weights in the form of $(y_{i-1},x_i)$ and $(y_{i-1},z_i)$ have 0 uncorrelated cost, for $1 \leq i <n$. Consequently, based on these link weight assignments, the MC-NSRLG problem in Fig.~\ref{Fig: NPHardPairCNC} can be reduced to the SP-SRLG problem in Fig.~\ref{Fig: NPHardCorrelated}, similar to the proof of the MCDCM problem in Section~\ref{Sec:CorreCut}.
 
However, we found that the MC-NSRLG problem is solvable in polynomial time when (1) only the nodal links in the form of $(u,v)$ and $(v,y)$ follow the SRLG-like correlated model and/or (2) for any node $u \in \mathcal{N}$, at most two nodal links $(u,v)$ and $(u,x)$ follow the SRLG-like correlated model. To prove case (1), let us first study the following theorem: 

\begin{theorem} \label{Theo:NodalCut}
Any two links in the form of $(u,v)$ and $(v,y)$ will never both appear in the optimal solution of the MC-NSRLG problem in case (1).
\end{theorem}

\begin{IEEEproof}
Suppose $s$ and $t$ are separated by a min-cut $C$ such that $s$ is in the subset $A$ and $t$ is in the subset $B$. A proof by contradiction: we assume $(u,v)$ and $(v,y)$ are both in the min-cut $C$. Since $C$ is the min-cut that separates $s$ and $t$, then node $u$ should be in subset $A$, otherwise if node $u$ is in subset $B$, there is no need to use $(u,v)$ and $(v,y)$ as the cut links, since their existence does not affect the connectedness between $A$ and $B$. Similarly, node $v$ is in subset $B$, otherwise if node $v$ is also in $A$, there is no need to cut link $(u,v)$. Based on this analysis, if $y$ is in $A$, then $(v,y)$ is not necessarily the link in the cut $C$, since link $(v,y)$ does not affect the connectedness from $A$ to $B$. However, if node $y$ is in $B$, link $(v,y)$ is also not necessarily the link in the min-cut, since nodes $v$ and $y$ are in the same subsets, which results in a contradiction.
\end{IEEEproof}

Based on Theorem \ref{Theo:NodalCut}, we can use conventional Linear Programming (LP) for solving the MC-NSRLG problem under case (1). Following \cite{Tamir94}, the LP is as follows.

\textbf{Objective:}%
\begin{equation}
\label{Eq: MinCutObj1}
\min\text{ } \sum\limits_{(u,v)\in\mathcal{L}'} c(u,v) \cdot h_{u,v} %
\end{equation}

\textbf{Constraints:}%

\begin{equation}
\label{Eq: MinCut1}
h_{s,t} \geq 1
\end{equation}

\begin{equation}
\label{Eq: MinCut2}
h_{u,v}+h_{v,y} \geq h_{u,y}, \text{  } \forall  u,v,y \in \mathcal{N}: u \neq v \neq y
\end{equation}

\begin{equation}
\label{Eq: MinCut3}
h_{u,v} \geq 0, \text{ } \forall u,v \in \mathcal{N}: u \neq v
\end{equation}
where $c(u,v)$ stands for the link weight (i.e., capacity) of link $(u,v)$ in the deterministic network and $h_{u,v}$ is an indicator denoting whether $(u,v)$ belongs to the cut.

\begin{figure}[tbh]
\centering
\includegraphics[trim = 0mm 0mm 0mm 0mm,clip,width=0.45\textwidth]{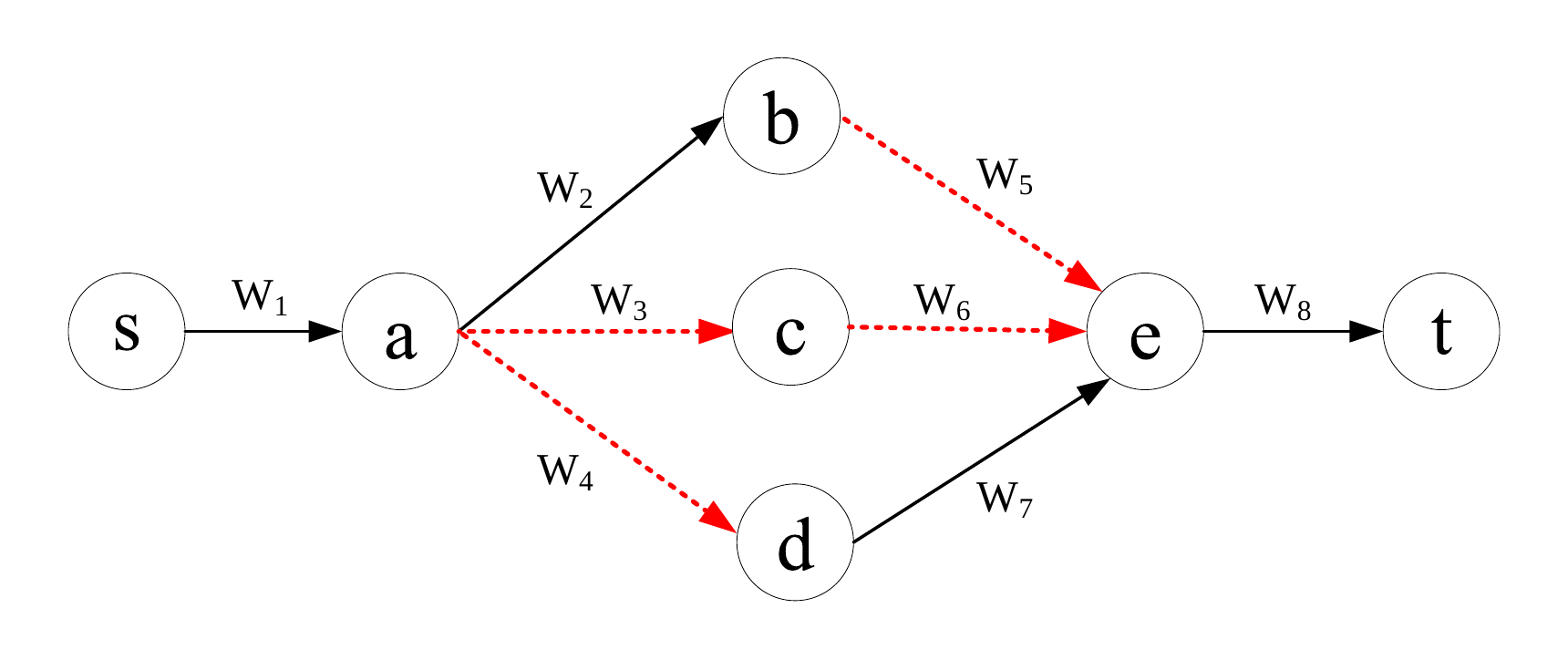}
\caption{An example network with dotted links following the SRLG-like correlated model.}%
\label{Fig:AuxExpCut}%
\end{figure}

\begin{figure}[tbh]
\centering
\includegraphics[trim = 0mm 0mm 0mm 0mm,clip,width=0.5 \textwidth]{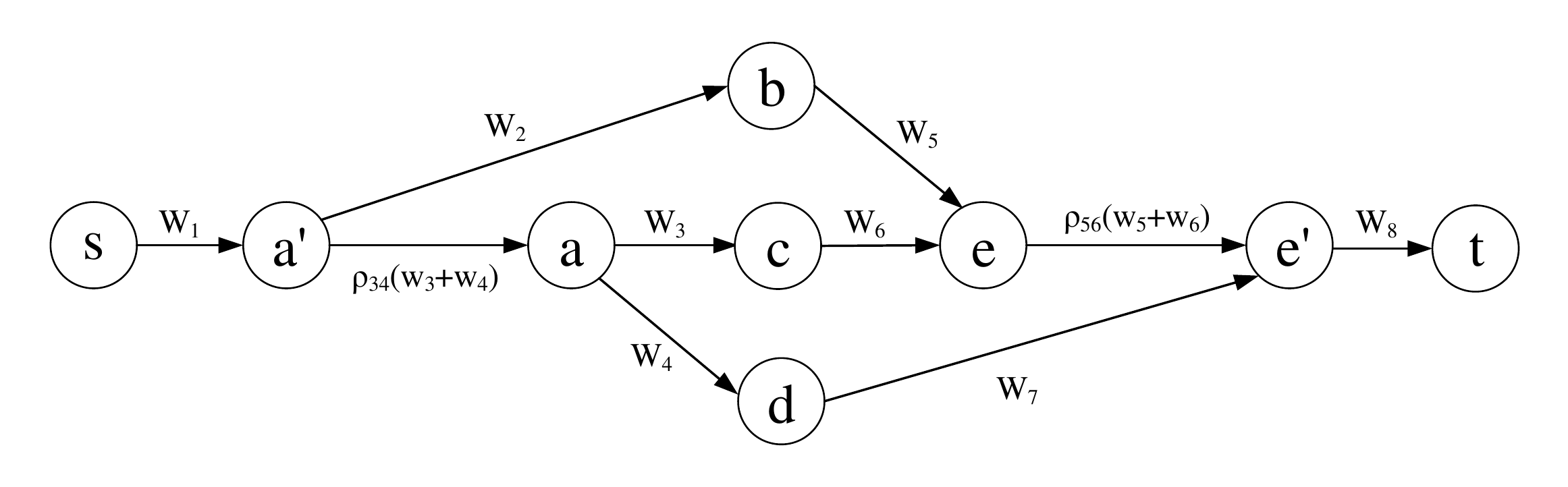}
\caption{Auxiliary graph of Fig.~\ref{Fig:AuxExpCut} for the MC-NSRLG problem.}%
\label{Fig:AuxGrapCut}%
\end{figure}

Under case (2), the MC-NSRLG problem can be solved in polynomial time by running the above LP (Eqs.~(\ref{Eq: MinCutObj1})-(\ref{Eq: MinCut3})) on an auxiliary graph $G^{U}(\mathcal{N}^U,\mathcal{L}^U)$. The auxiliary graph can be derived from the original graph $G$ as follows:
\begin{enumerate}
\item For each pair of two links $(u,v) \in \mathcal{L}$ and $(u,x) \in \mathcal{L}$ that are correlated in $G$, create a new node $u'$, and draw link $(u',u)$ with weight $\rho_{vx} (w(u,v)+w(u,x))$ to represent the joint cost of links $(u,v)$ and $(u,x)$, where $\rho_{vx}$ represents the correlation coefficient between $(u,v)$ and $(u,x)$.

\item For any link $(a,u) \in \mathcal{L} $ and $(u,b) \in \mathcal{L} \backslash \{v,x\} $ such that $(u,v)$ and $(u,x)$ are correlated in $G$, draw link $(a,u')$ and $(u',b)$ in $G^{U}$ with weights $w(a,u)$ and $w(u,b)$, respectively. 

\item For each pair of two links $(v,u) \in \mathcal{L}$ and $(x,u) \in \mathcal{L}$ that are correlated in $G$, create a new node $u'$, and draw link $(u,u')$ with weight $\rho_{vx} (w(v,u)+w(x,u))$ to represent the total cost of links $(v,u)$ and $(x,u)$, where $\rho_{vx}$ represents the correlation coefficient between $(v,u)$ and $(x,u)$.

\item For any link $(a,u) \in \mathcal{L} \backslash \{v,x\} $ and $(u,b)\in \mathcal{L} $ such that $(v,u)$ and $(x,u)$ are correlated in $G$, draw link $(a,u')$ and $(u',b)$ in $G^{U}$ with weights $w(a,u)$ and $w(u,b)$, respectively. 

\item For the other links $(c,d) \in \mathcal{L}$, create link $(c,d)$ also in $G^U$ with the same weight. 

\end{enumerate}
The proposed auxiliary graph shares similarities with the auxiliary graph in Fig.~\ref{Fig:AuxGrap}. For example, Fig.~\ref{Fig:AuxGrapCut} is an auxiliary graph of the original graph shown in Fig.~\ref{Fig:AuxExpCut}. Moreover, we mention that the proposed auxiliary graph also applies when there are $m>2$ links starting from the same node that follow the SRLG-like correlated model, but the failure of one correlated link will trigger the other $m-1$ links to simultaneously ``fail'' (e.g., SRLG networks, inter-dependent networks). 

\subsection{Min-Cut under the Stochastic Correlated Model}

\begin{definition}
The Min-Cut under Stochastic Correlated Model (MCSCM) problem: In a given directed graph $\mathcal{G}(\mathcal{N},\mathcal{L})$ with link costs following the stochastic correlated model, the MCSCM problem is to find a cut $\mathcal{C}$ which partitions $G$ into two disjoint subsets $X$ ($X \in \mathcal{N}$) and $\mathcal{N}-X$ such that:
\begin{itemize}
\item $s$ and $t$ are in different subsets. 
\item the allocated cost of the cut $C$ is minimum. 
\item the total probability of realizing the cost value is no less than $P_c$.
\end{itemize}

\end{definition}

We propose a corresponding convex optimization formulation based on Eq.~(\ref{Eq: MinCutObj1})-Eq.~(\ref{Eq: MinCut3}):

\textbf{Objective:}%
\begin{equation}
\label{Eq: MinCutObj}
\min\text{ } \sum\limits_{(u,v)\in\mathcal{L}} x(u,v) \cdot h_{u,v} %
\end{equation}

\textbf{Constraints:}%

\begin{equation}
\label{Eq:CutProb}
\sum_{i \in \Omega} -\log\left( CDF_i(x(l^i_1),x(l^i_2),...,x(l^i_{m_i}))\right) \leq - \log (P_c)
\end{equation}

\begin{equation}
0 \leq x(u,v) \leq c^{\max}_{(u,v)} \text{ } \text{ } \forall (u,v) \in \mathcal{L}
\label{Eq:MinCutBoundConstr}%
\end{equation}

\begin{equation}
h_{s,t} \geq 1
\label{Eq:MinCutConstr1}%
\end{equation}

\begin{equation}
h_{u,v}+h_{v,y} \geq h_{u,y}, \text{  } \forall  u,v,y \in \mathcal{N}: u \neq v \neq y
\label{Eq:MinCutConstr2}%
\end{equation}

\begin{equation}
h_{u,v} \geq 0 , \text{ } \text{  } \forall  u,v \in \mathcal{N}: u \neq v 
\label{Eq:MinCutConstr3}%
\end{equation}
where $x(u,v),x(l^i_1),x(l^i_2),...,x(l^i_{m_i})$ indicate the allocated possible cost by links $(u,v)$, $l^i_1$, $l^i_2$,..., $l^i_{m_i}$, respectively. In particular, Eq. (\ref{Eq:CutProb}) ensures that the probability of realizing the min-cut cost is no less than $P_c$. 
More specifically, for each MCG we apply the multi-dimensional CDF functions to calculate the probability of realizing a cost. Since the CDF function is log-concave, $-\log\left( CDF_i(x(l^i_1),x(l^i_2),...,x(l^i_{m_i}))\right)$ is convex, and by summing all the MCGs together, it remains convex, which indicates that Eq. (\ref{Eq:CutProb}) is convex.

It remains to show that Eq. (\ref{Eq: MinCutObj}) is convex. In general, the product of two convex functions is not always convex, however, according to~\cite[pp. 119]{boyd04}, one special case is: ``If functions $f$ and $g$ are convex, both nondecreasing (or nonincreasing), and positive (nonnegative) functions on an interval, then $f \cdot g$ is convex.'' Therefore, for each $(u,v)\in\mathcal{L}$, $x(u,v) \cdot h_{u,v}$ is convex.
\section{Related Work} \label{Sec:CorreRelWork}
\subsection{Routing with correlated link weights}
In a network with each link having multiple additive link weight metrics (e.g., delay, cost, jitter, etc.), the Quality of Service (QoS) routing problem is to find a path that satisfies a given constraints vector. Kuipers and Van Mieghem \cite{Kuipers03} study the QoS routing problem under correlated link weights. 
Another common source of correlation is Shared-Risk Link Groups (SRLGs). Sometimes one SRLG can also be represented by one color, but they share the same meaning in terms of reliability. In this context, Yuan \emph{et al.} \cite{Yuan05} prove that the Minimum Color Single-Path (MCSiP) problem is NP-hard. Yuan \emph{et al.} also prove that finding two link-disjoint paths with total minimum distinct amount of colors or least amount of coupled/overlapped colors is NP-hard. Lee \emph{et al.} \cite{Lee10} propose a probabilistic SRLG framework to model correlated link failures and develop an Integer Nonlinear Programming (INLP) formulation to find one unprotected path or two link-disjoint paths with the lowest failure probability.

There is also some literature dealing with correlated routing problems in stochastic networks \cite{commag2014}. For example, in \cite{Waller02} only two possible states are assumed, which are congested and uncongested, and each state corresponds to a cost value. A probability matrix $P^{u,v,y}_{a,b}$, which represents the probability that if $(u,v)$ is in state $a$ then $(v,y)$ is in state $b$, is given. Two similar link weight models, called link-based congestion model and node-based congestion model, are proposed in \cite{Fan05B}. Based on these models, \cite{Waller02, Fan05B} define and solve the least expected routing problem, which is to find a path from the source to the destination with minimum expected costs. However, in \cite{Waller02, Fan05B} there are only two possible states for each link and only the correlation of the adjacent links is known. We assume a more general (and different) stochastic correlated model, where as long as the links (not necessarily adjacent) are correlated, their joint CDF for allocating costs is known.   

\subsection{Min-Cut in Conventional Networks}
The $(s,t)$ Min-Cut problem refers to partitioning the network into two disjoint subsets, such that nodes $s$ and $t$ are in different subsets and the total weight of the cut links is minimized. This problem can be solved by finding the maximum flow from $s$ to $t$ \cite{Ford56}. There is also a lot of work on the Min-Cut problem with no specified node pairs $(s,t)$. A summary and comparison of polynomial-time algorithms to solve the Min-Cut problem can be found in \cite{Chekuri97}. The fastest algorithm to solve the Min-Cut problem has a time complexity of $O(L \log ^3 N)$ and was proposed by Karger \cite{Karger00}. Accordingly, the Min-Cut problem can be tackled by solving at most $N-1$ times the $(s,t)$ Min-Cut problem.

\subsection{Constrained Maximum Flow}
As a dual of the min-cut problem, the maximum flow problem in conventional networks is solvable in polynomial time \cite{Ford56}. However, this problem becomes NP-hard if some constraints are imposed on the links. Suppose that negative disjunctive constraints indicate that a certain set of links cannot be used simultaneously for the optimal solution, while positive disjunctive constraints force at least one of a certain set of links to be present in the optimal solution. Pferschy and Schauer \cite{Pferschy13} prove that the maximum flow problems with both negative and positive disjunctive constraints are NP-hard and do not admit a Polynomial-Time Approximation Scheme (PTAS). For example, the disjunctive constraint corresponds to the correlated link weights, so the maximum flow problem in correlated networks is also NP-hard and does not admit a PTAS. Assuming the link's bandwidth and delay follow a log-concave distribution, Kuipers \emph{et al.} \cite{ICNP14} propose a polynomial-time convex optimization formulation to find the maximum flow in so-called stochastic networks. When a delay constraint is imposed on each path, the maximum flow problem is NP-hard. To solve it, Kuipers \emph{et al.} \cite{ICNP14} propose an approximation algorithm and a tunable heuristic algorithm.

\section{Conclusion} \label{Sec:CorreConclusion}
In this paper, we have studied the shortest path problem and the min-cut problem in correlated networks under two link-weight models, namely (i) the deterministic correlated model and (ii) the (log-concave) stochastic correlated model. We have proved that these two problems are NP-hard under the deterministic correlated model, and cannot be approximated to arbitrary degree, unless P=NP. Subsequently, we have proposed exact algorithms to solve them. In particular, we have shown that both of them are solvable in polynomial time under a (constrained) nodal deterministic correlated model. For the stochastic correlated model, we have shown that these two problems can be solved by convex optimization.

\bibliographystyle{IEEEtran}

\end{document}